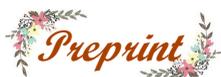

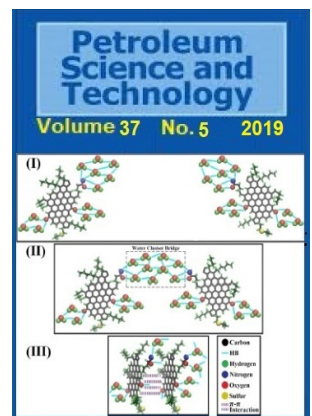

# Microscopic details of asphaltenes aggregation onset during waterflooding

**Salah Yaseen**[a] and **G.Ali Mansoori**[b]

[a]Dep't of Chemical Engineering, Univ. of Illinois at Chicago, Chicago, Illinois, USA; *syasee3@uic.edu*
[b]Dep'ts of Bio- & Chemical Engineering, Univ. of Illinois at Chicago, Chicago, Illinois, USA; *mansoori@uic.edu*

**ABSTRACT**

We report detailed microscopic studies of asphaltenes aggregation onset during waterflooding of petroleum reservoirs. To achieve this objective, a series of simulations are performed on asphaltenic-oil miscibilized with water at high pressure and temperature through molecular dynamics. Results of this simulation onset are applicable to asphaltenes behavior in real crude oils. Our simulation results illustrate that the aggregation onset in waterflooding generally follows three sequential steps: (i) Asphaltene-water interaction; (ii) Water bridging; (iii) Face-to-face stacking. Then, asphaltene-water and water-water hydrogen-bonding network surround every aggregate boosting the intensity of aggregation onset. We intend to utilize such understanding of these details in our predictive and preventive measures of arterial blockage in oil reservoirs during waterflooding.

## 1. Introduction

To practicing engineers, the importance of knowledge about the behavior of asphaltenes in petroleum is similar to the knowledge needed by cardiologists regarding cholesterol in the arteries of patients (Leontaritis and Mansoori 1988). Asphaltenes, if present in a petroleum fluid, tend to aggregate and deposit in flow paths causing plugging in the reservoir porous media and transmission pipelines (Escobedo and Mansoori 1997; Vazquez and Mansoori, 2000; Branco et al. 2001; Hu et al. 2004; Mansoori, 2009).

There has been a lack of an understanding of the nature of asphaltenes instability during waterflooding. Recently, we reported a series of special molecular dynamics (MD) studies on the onset of asphaltenes aggregation during a waterflooding process (Yaseen and Mansoori 2017; 2018a; 2018b). Results of our simulation onset are applicable to asphaltenes behavior in real crude oils. According to our studies, asphaltenes became poorly soluble when water was miscibilized in the oil phase at high pressures and temperatures. Face-to-face contact was the predominant stacking of the onset of asphaltenes aggregation. Additionally, we showed that using high-salinity brines (25 wt.%), instead of pure-water, significantly reduced the intensity of asphaltenes aggregation at the onset.

The microscopic details of asphaltenes aggregation onset in waterflooding of reservoirs remained uncertain. Motivated by our previous studies, considerable attention is given to understanding the reasons behind asphaltenes aggregation onset due to waterflooding. Specifically, we study the role of various segments of seven different asphaltene molecules in this research. For this purpose, we perform an MD study on asphaltenic-oil/pure-water systems under a reservoir

**Corresponding author:** Salah Yaseen; *syasee3@uic.edu; salahyaseen1983@gmail.com*

condition. We choose ortho-xylene as the oil medium because it is the best hydrocarbon solvent for asphaltenes. The onset data produced in this study is also valid for the real asphaltenic crude oils. This study will provide, from a molecular level, an advanced insight into the relation between water-injection and asphaltenes aggregation onset at reservoir conditions.

## 2. Molecular asphaltenes

Seven different asphaltene molecules labeled as $A_1$, $A_2$, … $A_7$, are used in this study as shown in Figure 1. These molecules were also used in our previous studies (Yaseen and Mansoori 2017; 2018a; 2018b). Their molecular weights are ranged between 543 ($A_1$) and 1197 ($A_4$). Their molecular architectures are classified into the island ($A_1$, $A_3$, $A_4$, $A_5$, $A_7$) and archipelago ($A_2$, $A_6$). Their elemental compositions are mainly composed of carbon and hydrogen atoms. However, their structures also contain heteroatoms, including nitrogen (N), oxygen (O), and sulfur (S).

Presence of N- and O-segments in asphaltenes structures are responsible for possible hydrogen-bonding. Such segments include amide, amine, carboxyl, hydroxyl, and pyridyl. On the other hand, S-segments, such as sulfide, thioanisole, and thiophene, could not form hydrogen-bonds.

To investigate the role of heteroatom-segments on the onset of asphaltenes aggregation, two other "hypothetical" groups of seven molecules are considered in our study. The first "hypothetical" group (Group B) are assumed to have the same structures as the asphaltenes in Figure 1, but their S-segments are removed and replaced with carbon and hydrogen atoms (see Figure 2). The second "hypothetical" group (Group C) are assumed to have the same structures as the asphaltenes in Figure 1, but their N and O atoms are removed and replaced with carbon and hydrogen atoms as shown in Figure 2.

In what follows through MD simulation of asphaltene and the two hypothetical sets of molecules we report the microscopic reasons for the onset of asphaltenes aggregation during petroleum reservoir waterflooding.

## 3. MD simulation method used in this study

MD simulation is a powerful tool to gain certain insight into the microscopic (atomic level) characteristics of materials. The trajectories of atoms are determined by solving Newton's equation of

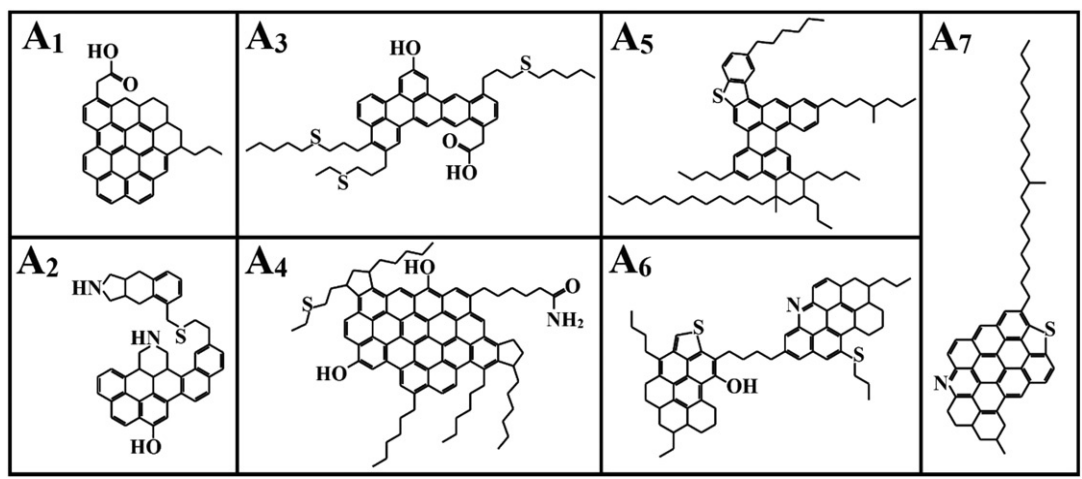

**Figure 1.** Molecular structures of asphaltene molecules used in this study.

# (i) Hypothetical B Molecules

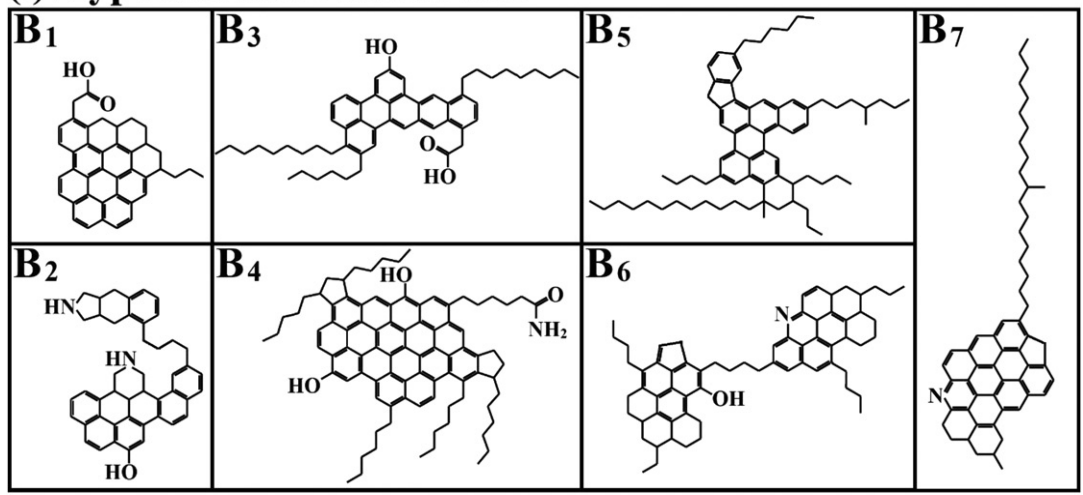

# (ii) Hypothetical C Molecules

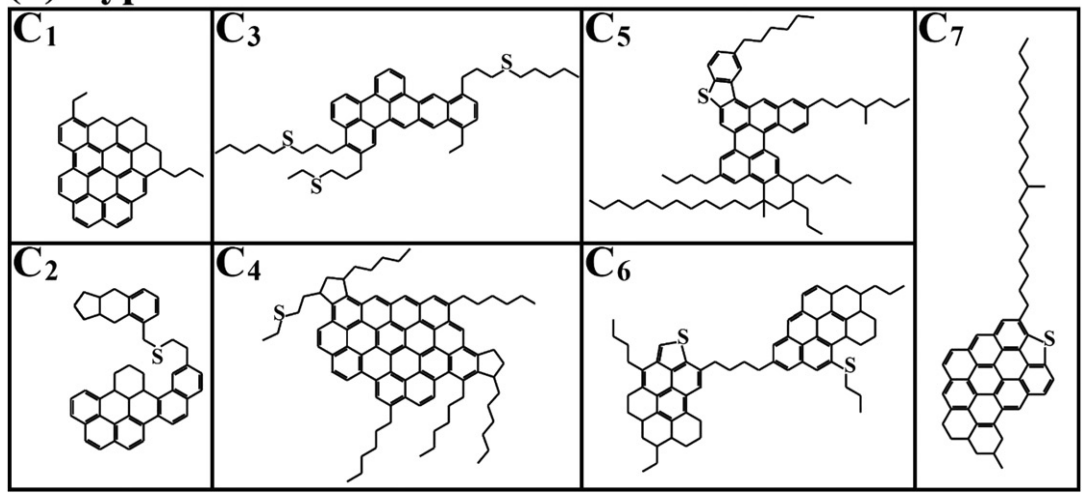

**Figure 2.** (i) Hypothetical B molecules, having the same structures as asphaltenes in Figure 1, but lacking sulfur in their structure. (ii) Hypothetical C molecules, having the same structures as asphaltenes in Figure 1, but lacking nitrogen and oxygen in their structures.

motion. MD simulation has been found to be quite useful for investigating the aggregation onset of asphaltenes in various model petroleum fluids (Mohammed and Mansoori 2018; Khalaf and Mansoori 2018a; 2018b). In the current study, GROMACS 5.1.2 software package (www.gromacs. org) is employed to perform MD simulations. The bonding and non-bonding interactions between the neighboring atoms are calculated by force fields. In this report, water is modeled by the SPC/E potential (Berendsen, Grigera, and Straatsma 1987). On the other hand, asphaltenes and ortho-xylene are represented by OPLS-AA potential (Jorgensen, Maxwell, and Tirado-Rives 1996).

During a typical MD simulation, the simulation box is constructed such that it is composed of two phases: the oil phase and the aqueous phase. The oil phase is selected to be composed of ortho-xylene and asphaltenes. Twenty-four asphaltene molecules (10 wt.%) are packed into the oil phase, and the simulation box is then expanded in the Y-direction and packed with pure-water.

Once the simulation box is constructed, it undergoes an energy minimization step to relax the system towards equilibrium. MD simulations are performed with periodic boundary conditions in X-, Y-, and Z-directions. They are performed for 20 ns duration in the NPT ensemble at constant temperature (550 K) and constant pressure (200 bar). The detailed description of system preparation and the MD method are reported in our previous publication (Yaseen and Mansoori 2017).

## 4. Observations resulting from MD simulations

### 4.1. Effect of heteroatom-segments on aggregation intensity

We calculate radial distribution functions (RDFs) of asphaltenes and the two sets of hypothetical molecules (see Figure 3). The intensity of aggregation onsets is proportional to heights of RDF first peaks. According to Figure 3, asphaltenes and hypothetical B molecules have similar first peak heights. However, RDFs of hypothetical C molecules possess little or no peaks, indicating a lack of aggregation. Accordingly, N- and O-segments are the only causes of asphaltenes aggregation onsets in waterflooding.

According to RDFs in Figure 3, separation distances of peaks are ranged between 0.38 and 0.50 nm, which are within the face-to-face cutoff range (Yaseen and Mansoori 2018a).

### 4.2. Distribution of water near heteroatom-segments

According to RDFs in Figure 4, S-segments possess little or no peaks. This indicates that water does not tend to concentrate near these segments. On the contrary, every N- and O-segment possess at least one remarkable peak. They reveal that water highly occupies spaces near these segments.

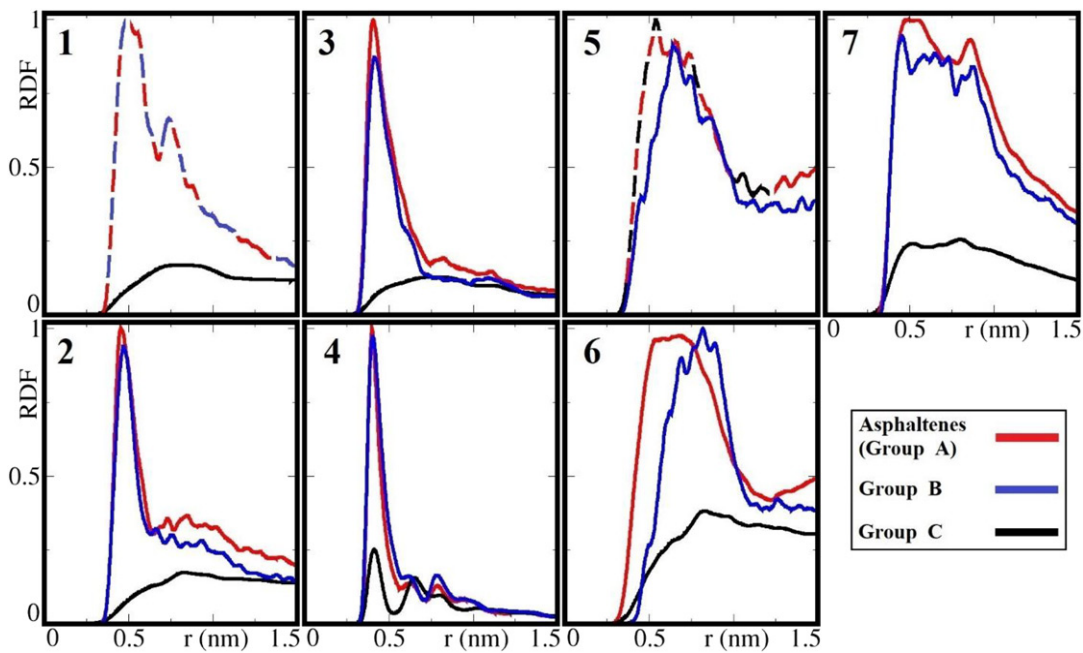

**Figure 3.** RDFs of asphaltenes-asphaltenes stacking as compared with RDFs of hypothetical molecules (B and C). Differences of RDFs of asphaltenes with hypothetical molecules, B & C, are indications of the roles of heteroatoms (N, O, S) in asphaltenes aggregation. All the calculations are for the centers of mass interactions. The two-colored graphs in Figures 3-1 and 3-5 indicate the similarity of results for asphaltenes with B and C molecules, respectively.

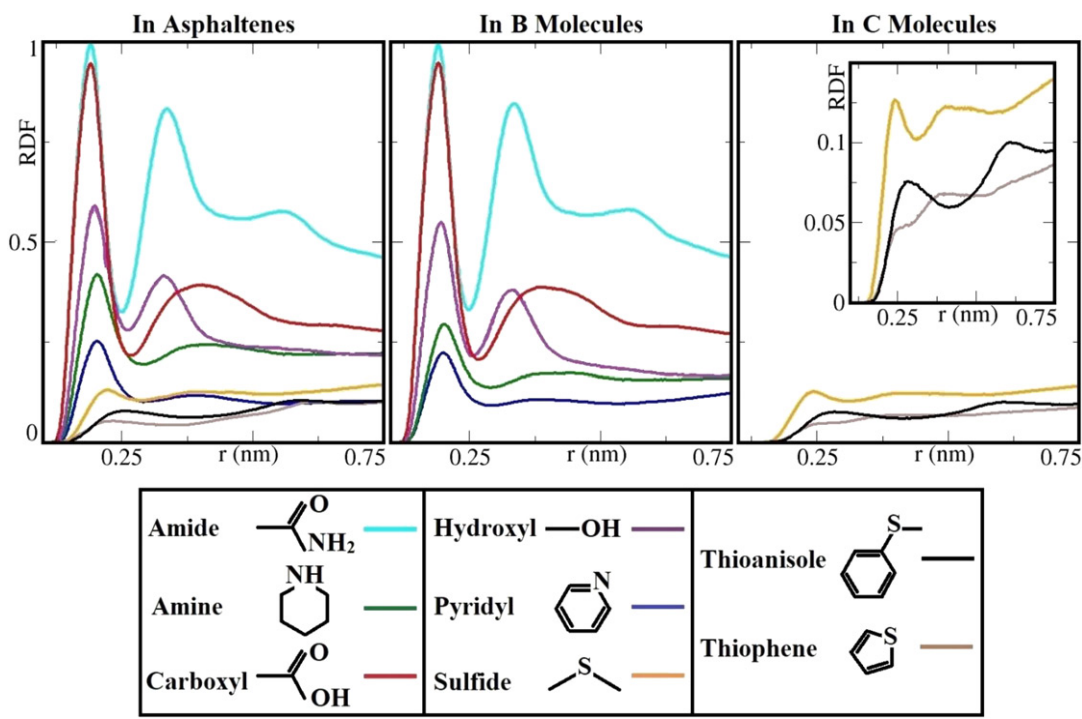

**Figure 4.** Relative distributions of water around various segments listed above (represented by RDFs) appearing in asphaltenes, B and C molecules.

**Table 1.** The average number of hydrogen-bonds of asphaltenes and hypothetical molecules at 550 K and 200 bar

| Hydrogen-bonds | $i=1$ | $i=2$ | $i=3$ | $i=4$ | $i=5$ | $i=6$ | $i=7$ |
|---|---|---|---|---|---|---|---|
| $HB_{Ai\text{-}Ai}$ | 1.1 | 1.0 | 2.9 | 5.0 | 0 | 0.9 | 0 |
| $HB_{Ai\text{-}W}$ | 28.7 | 29.2 | 47.1 | 81.8 | 0 | 17.8 | 4.9 |
| $HB_{Bi\text{-}Bi}$ | 0.8 | 0.9 | 3.2 | 5.0 | 0 | 1.1 | 0 |
| $HB_{Bi\text{-}W}$ | 29.0 | 29.2 | 46.5 | 81.2 | 0 | 18.0 | 4.8 |
| $HB_{Ci\text{-}Ci}$ | 0 | 0 | 0 | 0 | 0 | 0 | 0 |
| $HB_{Ci\text{-}W}$ | 0 | 0 | 0 | 0 | 0 | 0 | 0 |

## 4.3. Hydrogen-Bonding of asphaltenes

We calculate average numbers of hydrogen-bonds of asphaltenes ($HB_{Ai\text{-}Ai}$, $HB_{Ai\text{-}W}$) and hypothetical molecules ($HB_{Bi\text{-}Bi}$, $HB_{Bi\text{-}W}$, $HB_{Ci\text{-}Ci}$, $HB_{Ci\text{-}W}$), using the *gmx hbond* analysis tool (www.gromacs.org). According to Table 1, the numbers of $HB_{Ai\text{-}Ai}$ and $HB_{Bi\text{-}Bi}$ are rather small as compared with $HB_{Ai\text{-}W}$ and $HB_{Bi\text{-}W}$, respectively. In the case of hypothetical C molecules, both $HB_{Ci\text{-}Ci}$ and $HB_{Ci\text{-}W}$ are zero.

## 4.4. Distribution of N & O heteroatoms at the onset of asphaltene aggregation

According to Figure 5, RDFs of asphaltenes and hypothetical B molecules have apparent peaks in such a separation distance ranging between 0.33 nm and 0.38 nm. These peaks reveal that the N- and O-segments are close to each other in the stacked asphaltenes and hypothetical B molecules.

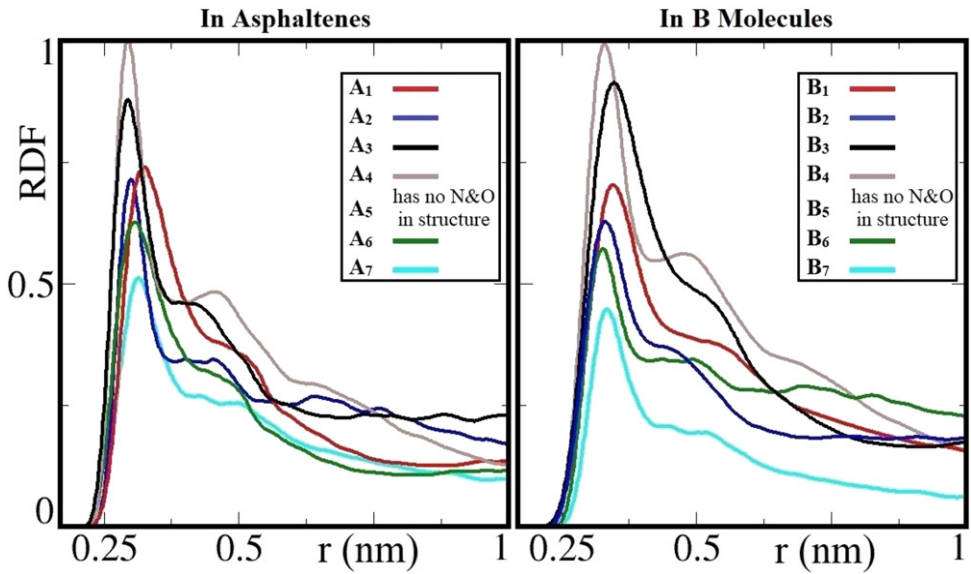

**Figure 5.** Relative distributions of N- and O-segments in the stacked asphaltenes and hypothetical B molecules at their onset of aggregation. The distribution is represented by segment-segment RDFs of one molecule with respect to another.

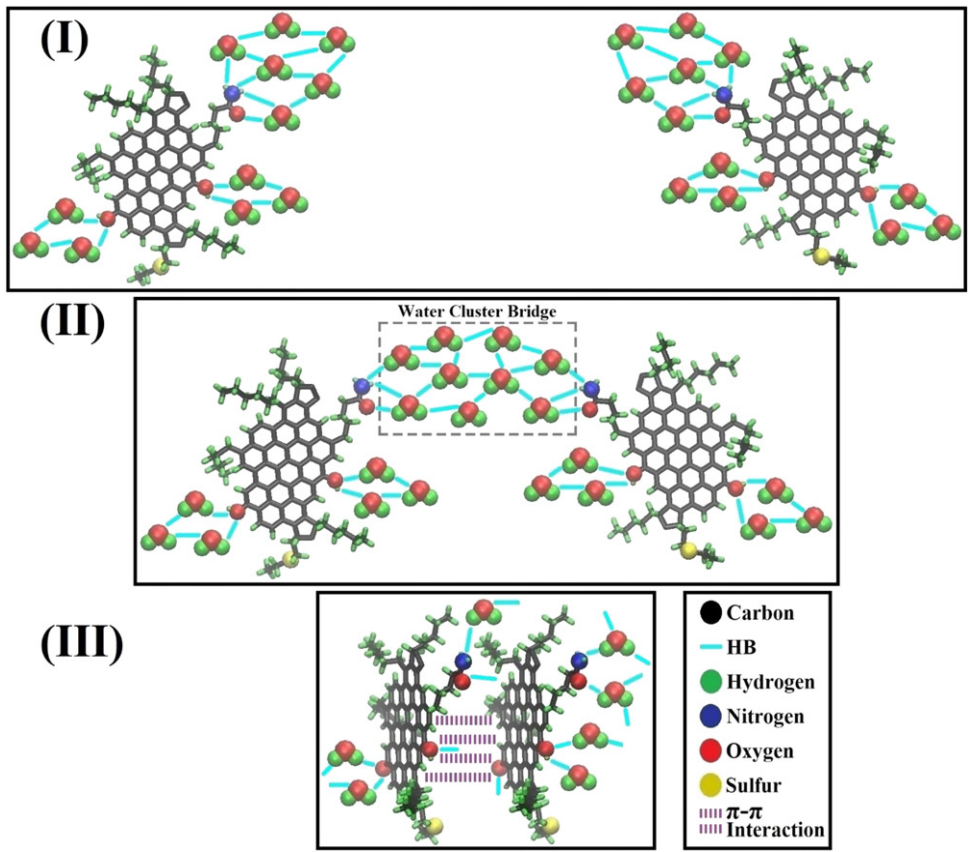

**Figure 6.** Schematic representation of sequential steps involved in asphaltenes aggregation due to waterflooding. In these figures, oil is omitted for clarity. Stage I: Asphaltene-water interaction via $HB_{A_i-W}$ Stage II: Formation of water bridges by $HB_{W-W}$ Stage III: Face-to-face stacking configuration of asphaltenes via $\pi-\pi$ interaction.

## 5. Discussion and conclusions

In this report, a series of MD simulations are performed to understand why and how the onset of asphaltenes aggregation occurs. We employ simulations on asphaltenic-oil/pure-water systems under 550 K-200 bar. MD results illustrate that the association of asphaltenes is due to the presence of the N- and O-segments in their structures. We demonstrate the fact that S-segments have little or no roles in asphaltenes aggregation onset due to waterflooding. According to our special MD simulations reported in this paper, aggregation onset of asphaltenes follow three consecutive stages (see Figure 6):

- Stage I: Once water molecules diffuse into the oil phase, they migrate toward the N- and O-segments of asphaltenes and form hydrogen bonds until all N- and O-segments are bounded by clusters of water molecules.
- Stage II: Hydrogen bonds are formed between water clusters surrounding asphaltene molecules. Then, water clusters serve as bridges between every pair of neighboring water-covered asphaltenes.
- Stage III: Once two water-covered asphaltenes are close enough to each other, $\pi$-$\pi$ interaction occurs between aromatic moieties, and a face-to-face asphaltenes dimer is formed. As a result, $\pi$-$\pi$ interaction, and hydrogen-bonds network maintain the stability of asphaltene aggregates.

We are aware of the fact that MD simulation is incapable of predicting the behavior of complex petroleum fluids. While equations of state, in principle, are capable of predicting the behavior of lighter (up to seven-carbon atom hydrocarbons), one needs to use a combination of polydisperse polymer and colloidal solution theories along with kinetic theories of aggregation and association in order to deal with the behavior of asphaltenic oils (Mansoori, 2009). What we have reported here is the onset of interaction/aggregation of two individual dissolved-in-oil asphaltene molecules as a result of the presence of miscible water in oil. We are confident this result is applicable to asphaltenes behavior in real crude oils.

### Acknowledgments

Salah Yaseen is grateful to the Higher Committee for Education Development in Iraq (HCED) for his financial support during this research.

### ORCID

Salah Yaseen  https://orcid.org/0000-0003-0724-8044
G.Ali Mansoori  https://orcid.org/0000-0003-3497-8720

### References

Branco, V. A. M., G. A. Mansoori, L. C. D. A. Xavier, S. J. Park, and H. Manafi. 2001. Asphaltenes flocculation and collapse from petroleum fluids. *Journal of Petroleum Science and Engineering* 32 (2–4):217–30. doi: 10.1016/S0920-4105(01)00163-2.

Berendsen, H. J. C., J. R. Grigera, and T. P. Straatsma. 1987. The missing term in effective pair potentials. *The Journal of Physical Chemistry* 91 (24):6269–71. doi: 1021/j100308a038.

Escobedo, J., and G. A. Mansoori. 1997. Viscometric principles of onsets of colloidal asphaltene flocculation in paraffinic oils and asphaltene micellization in aromatics. *SPE Production & Facilities* 12 (02):116–22. doi: 10.2118/28729-PA.

Hu, Y. F., S. Li, N. Liu, Y. P. Chu, S. J. Park, G. A. Mansoori, and T. M. Guo. 2004. Measurement and corresponding states modeling of asphaltene precipitation in Jilin reservoir oils. *Journal of Petroleum Science and Engineering* 41 (1–3):169–82. doi: 10.1016/S0920-4105(03)00151-7.

Jorgensen, W. L., D. S. Maxwell, and J. Tirado-Rives. 1996. Development and testing of the OPLS all-atom force field on conformational energetics and properties of organic liquids. *Journal of the American Chemical Society* 118 (45):11225–36. doi: 10.1021/ja9621760.

Khalaf, M. H., and G. A. Mansoori. 2018a. Asphaltenes aggregation during petroleum reservoir air and nitrogen flooding. *Journal of Petroleum Science and Engineering* 173: 1121–9. doi:10.1016/j.petrol.2018.10.037.

Khalaf, M. H., and G. A. Mansoori. 2018b. A new insight into asphaltenes aggregation onset at molecular level in G.A. crude oil (an MD simulation study). *Journal of Petroleum Science and Engineering* 162:244–50. doi: 10.1016/j.petrol.2017.12.045.

Leontaritis, K. J., and G. A. Mansoori. 1988. Asphaltene deposition: A survey of field experiences and research approaches. *Journal of Petroleum Science and Engineering* 1 (3):229–39. doi: 10.1016/0920-4105(88)90013-7.

Mansoori, G. A. 2009. A unified perspective on the phase behaviour of petroleum fluids. *International Journal of Oil, Gas and Coal Technology* 2 (2):141–67. doi: 10.1504/IJOGCT.2009.024884.

Mohammed, S., and G. A. Mansoori. 2018. Effect of $CO_2$ on the interfacial and transport properties of water/binary and asphaltenic oils: Insights from molecular dynamics. *Energy & Fuels* 32 (4):5409–17. doi: 10.1021/acs.energyfuels.8b00488.

Vazquez, D., and G. A. Mansoori. 2000. Identification and measurement of petroleum precipitates. *Journal of Petroleum Science and Engineering* 26 (1–4):49–55. doi: 10.1016/S0920-4105(00)00020-6.

Yaseen, S., and G. A. Mansoori. 2017. Molecular dynamics studies of interaction between asphaltenes and solvents. *Journal of Petroleum Science and Engineering* 156 :118–24. doi: 10.1016/j.petrol.2017.05.018.

Yaseen, S., and G. A. Mansoori. 2018a. Asphaltenes aggregation due to waterflooding (A molecular dynamics simulation study). *Journal of Petroleum Science and Engineering* 170:177–83. doi:10.1016/j.petrol.2018.06.043.

Yaseen, S., and G. A. Mansoori. 2018b. Asphaltene aggregation onset during High-Salinity waterflooding of reservoirs (A molecular dynamic study). *Petroleum Science and Technology* 36:1725–32. doi: 10.1080/10916466.2018.1506809.

## Nomenclature, Abbreviations, Formulas

| | |
|---|---|
| $A_i$ | Notation for asphaltenes, where i = 1, 2 … , 7. |
| Amide | —$C^{alkyl}$—$ONH_2$ |
| Amine | —$C^{alkyl}$—NH—$C^{alkyl}$— |
| bar | A unit of pressure measurement, which is equal to 14.5 pounds per square inch (psi). |
| $B_i$ | Notation for hypothetical B molecules, where i = 1, 2 … , 7. |
| Carboxyl | —$C^{alkyl}$—OOH |
| $C_i$ | Notation for hypothetical C molecules, where i = 1, 2 … , 7. |
| $HB_{Ai\text{-}Ai}$ | Hydrogen-bonds among asphaltenes. |
| $HB_{Ai\text{-}W}$ | Hydrogen-bonds between asphaltenes and water. |
| $HB_{Bi\text{-}Bi}$ | Hydrogen-bonds among hypothetical B molecules. |
| $HB_{Bi\text{-}W}$ | Hydrogen-bonds between hypothetical B molecules and water. |
| $HB_{Ci\text{-}Ci}$ | Hydrogen-bonds among hypothetical C molecules. |
| $HB_{Ci\text{-}W}$ | Hydrogen-bonds between hypothetical C molecules and water. |
| $HB_{W\text{-}W}$ | Hydrogen-bonds among water. |
| Heteroatom | Term is used to describe non-hydrocarbon atoms, such as nitrogen (N), oxygen (O), sulfur (S) |
| Hydroxyl | —$C^{aromatic}$—OH |
| K | Degree Kelvin is a unit of temperature measurement. |
| MD | Molecular dynamics |
| N-segment | Nitrogen-containing segment |
| O-segment | Oxygen-containing segment |
| OPLS_AA | Optimized Potential for Liquid Simulations - All Atom force field |
| Pyridyl | —$C^{aromatic}$—N—$C^{aromatic}$— |
| RDF | Radial distribution function |
| S-segment | Sulfur-containing segment |
| SPC/E | Simple Point Charge/Extended force field |
| Sulfide | —$C^{alkyl}$—S—$C^{alkyl}$— |
| Thioanisole | —$C^{aromatic}$—S—$C^{alkyl}$— |
| Thiophene | —$C^{aromatic}$—S—$C^{aromatic}$— |
| $\pi\text{-}\pi$ | A type of non-covalent interaction that involves aromatic/benzene rings |